\documentclass{ptephy_v1}

\usepackage{graphics}
\usepackage{bm,latexsym,amsmath,amssymb,amsfonts,mathrsfs,textcomp}
\usepackage[colorlinks=true, pdfstartview=FitV, linkcolor=red, citecolor=blue, urlcolor=blue]{hyperref}
\usepackage[normalem]{ulem}

\newcommand\sect[1]{{\it #1.}---}

\newcommand{\diag}{\mathrm{diag}}

\newcommand{\home}{\hat{\omega}}
\newcommand{\eps}{\hbar}

\newcommand{\U}{\text{U}}

\newcommand{\bzero}{\bm{0}}

\newcommand{\diff}{\mathrm{d}}

\newcommand{\para}{\parallel}

\newcommand{\eqq}{~\Leftrightarrow~}

\begin{document}

\title{Gyrohydrodynamics: \\
Relativistic spinful fluid with strong vorticity}

\author[1]{Zheng Cao}
\affil[1]{Physics Department and Center for Particle Physics and Field Theory, Fudan University, Shanghai 200438, China}

\author[2,3,a]{Koichi Hattori}
\affil[2]{Zhejiang Institute of Modern Physics, Department of Physics, 
Zhejiang University, Hangzhou, Zhejiang 310027, China}
\affil[3]{Research Center for Nuclear Physics, Osaka University, 
10-1 Mihogaoka, Ibaraki, Osaka 567-0047, Japan}

\author[4,5,b]{Masaru Hongo}
\affil[4]{Department of Physics, Niigata University, Niigata 950-2181, Japan}
\affil[5]{RIKEN iTHEMS, RIKEN, Wako 351-0198, Japan}

\author[1,6,7,c]{Xu-Guang Huang}
\affil[6]{Key Laboratory of Nuclear Physics and Ion-beam Application (MOE), Fudan University, Shanghai 200433, China}
\affil[7]{Shanghai Research Center for Theoretical Nuclear Physics, NSFC and Fudan University, Shanghai 200438, China}

\author[5,d]{Hidetoshi Taya}

\affil[a]{{\tt koichi.hattori@zju.edu.cn}}
\affil[b]{{\tt hongo@phys.sc.niigata-u.ac.jp}}
\affil[c]{{\tt huangxuguang@fudan.edu.cn}}
\affil[d]{{\tt hidetoshi.taya@riken.jp}}

\begin{abstract} 
 We develop a relativistic (quasi-)hydrodynamic framework, dubbed the {\it gyrohydrodynamics}, to describe fluid dynamics of many-body systems with spin under strong vorticity based on entropy-current analysis. 
 This framework generalizes the recently-developed spin hydrodynamics to the regime where the spin density is at the leading order in derivatives but suppressed by another small parameter, the Planck constant $\hbar$, due to its quantum nature.
 Our analysis shows that the complete first-order constitutive relations of gyrohydrodynamics involve seventeen transport coefficients and are highly anisotropic. 
 \end{abstract}

\subjectindex{A52, D28, D31, E14, I55}

\maketitle

\sect{Introduction}
Hydrodynamics is a low-frequency and long-wavelength effective theory for interacting many-body systems~\cite{Landau:Fluid}. 
When incorporated with relativity, it has been applied very successfully in astrophysics, cosmology, and relativistic heavy-ion collisions. For example, relativistic hydrodynamics provides an important foundation for the understanding of the core-collapse supernova explosions, the structures and bulk properties of neutron stars and white dwarfs, the dynamics of accretion discs around compact objects, and the merging process of binary neutron stars and black holes; see, e.g., Refs.~\cite{Camenzind:book,Rezzolla:book,Baiotti:2016qnr,Paschalidis:2016vmz,Muller:2020ard}. In high-energy heavy-ion collisions, relativistic hydrodynamics has become the `standard model' to describe the evolution of quark-gluon plasma created during the collisions; see e.g., Refs.~\cite{Heinz:2013th,Florkowski:2017olj,Shen:2020mgh}. 

It is noteworthy that rotation of fluids, whether it is global or local (local rotation is characterized by fluid vorticity), plays a crucial role in all of the systems mentioned above.  One of the critical consequences of rotation is spin polarization when the constituents of fluids are spinful.  In particular, the strongest fluid vorticity has recently been detected in heavy-ion collisions~\cite{STAR:2017ckg}.  Measurements of hyperon polarization~\cite{STAR:2017ckg,Adam:2018ivw,Adam:2020pti} and vector-meson spin alignment~\cite{Acharya:2019vpe,STAR:2022fan,ALICE:2022sli} in heavy-ion collisions have indicated strong spin polarization of quarks and/or gluons in the quark-gluon plasma, supporting earlier theoretical ideas proposed in, e.g., Refs.~\cite{Liang:2004ph,Liang:2004xn,Betz:2007kg,Becattini:2007sr,Huang:2011ru,Becattini:2013fla}. As such, it is desirable to develop a relativistic hydrodynamic framework that deals appropriately with the spin degrees of freedom and the vorticity generation. Such a framework may pave the way to quantitative understanding of spin transports in quark-gluon plasma, supernova explosions, and so on. Specifically, it may be used to resolve the so-called sign problem of the local spin polarization, that is, the discrepancy between the experimental data for azimuthal-angle dependence of hyperon polarization and the theoretical calculations based on thermal vorticity (see below for definition); see reviews~\cite{Liu:2020ymh,Gao:2020vbh,Huang:2020xyr,Becattini:2020ngo,Becattini:2022zvf} and references therein for more details.

There have been great efforts on developing the framework of {\it spin hydrodynamics}~\cite{Montenegro:2017rbu,Montenegro:2017lvf,Florkowski:2017ruc,Florkowski:2018fap,Montenegro:2018bcf,Hattori:2019lfp,Li:2019qkf,Fukushima:2020ucl,Bhadury:2020puc,Montenegro:2020paq,Li:2020eon,Shi:2020htn,Garbiso:2020puw,Gallegos:2020otk,She:2021lhe,Hu:2021lnx,Gallegos:2021bzp,Peng:2021ago,Hongo:2021ona,Hu:2021pwh,Cartwright:2021qpp,Wang:2021wqq,Hongo:2022izs,Singh:2022ltu,Daher:2022xon,Gallegos:2022jow,Weickgenannt:2022zxs,Bhadury:2022qxd}. In most of these developments, the spin and vorticity are considered to be of similar magnitudes as the shear viscous correction to the hydrodynamics, meaning that they are assumed to be $O(\partial^1)$ in terms of the derivative expansion. 
This is possible, since vorticity is the skew gradient of velocity while the shear tenor is the symmetric gradient of velocity.  Thus, one expects that their strength may be comparable in magnitude. 
However, there is a significant difference between the two: the shear tensor vanishes at the global equilibrium, whereas vorticity does not.  This implies that, on the physical ground, these two can be regarded to be at different orders in derivatives in and near equilibrium (cf. Ref.~\cite{Li:2020eon} for another power counting scheme based on the same consideration). Furthermore, the discovery of strong spin polarization in the quark-gluon plasma also demands development of a hydrodynamic theory with high spin density and vorticity. 

Motivated by these observations, we will investigate a scenario in which vorticity is counted as zeroth order in derivatives, whereas other types of gradients of thermodynamic quantities 
are counted as first order in derivatives. 
The spin polarization induced by the zeroth-order vorticity is also regarded as the zeroth order in derivatives, but we assume it to be suppressed by another small parameter, i.e., the Planck constant $\hbar$ due to its quantum nature. 
We dub the resultant framework {\it gyrohydrodynamics}.

As we will demonstrate, the constitutive relations in gyrohydrodynamics have much richer structure than those in spin hydrodynamics~\cite{Hattori:2019lfp, Bhadury:2020puc, Fukushima:2020ucl, Gallegos:2021bzp, Hongo:2021ona, Hu:2021pwh, Gallegos:2022jow, Weickgenannt:2022zxs, Daher:2022xon}. 
Firstly, the leading-order constitutive relation describes the anisotropic pressure induced by strong vorticity.
Besides, we find seventeen anisotropic transport coefficients: three bulk, four shear, three rotational viscosities, four cross viscosities and three charge conductivities.
Among these transport coefficients, seven correspond to Hall-like transports, which do not induce the entropy production.
The appearance of the anisotropic transports is analogous to magnetohydrodynamics, but there are no Hall transports in magnetohydrodynamics in the strict hydrodynamic limit~\cite{Grozdanov:2016tdf,Hongo:2020qpv}.

We stress that spin is not a conserved quantity due to the inherent spin-orbit coupling in relativistic systems.  As a result, spin density is not a strict hydrodynamic variable. Therefore, spin hydrodynamics and thus gyrohydrodynamics should be regarded as {\it quasi}-hydrodynamics~\cite{Grozdanov:2018fic} or a hydro+ theory~\cite{Stephanov:2017ghc} valid in a regime where the relaxation time of spin density towards its local equilibrium value is much longer than that of other non-hydrodynamic modes~\cite{Hongo:2021ona}. To keep the notations simple, however, we will continue calling spin density a hydrodynamic variable and 
gyrohydrodynamics a hydrodynamic theory. 

Throughout the paper, we use the mostly-plus metric $\eta_{\mu\nu}= \diag (-1,+1,+1,+1)$, totally anti-symmetric tensor $\epsilon^{\mu\nu\rho\sigma}$ with $\epsilon^{0123}=+1$,
and notations $X^{(\mu\nu)} = \frac{1}{2}(X^{\mu\nu}+X^{\nu\mu})$ and $X^{[\mu\nu]} = \frac{1}{2}(X^{\mu\nu}-X^{\nu\mu})$ for an arbitrary tensor $X^{\mu\nu}$.

\sect{Entropy current analysis for rotating spinful fluids}
The hydrodynamic equations follow from Ward-Takahashi identities derived from symmetries of underlying microscopic theories.  In this Letter, we consider charged relativistic fluids that enjoy the translational, Lorentz, and vector $\U(1)$ symmetries.  These symmetries lead to the equations of motion, 
\begin{equation}
 \partial_\mu \Theta^\mu_{~\nu} = 0, \quad 
  \partial_\mu \Sigma^{\mu}_{~\nu\rho} 
  = - 2 \Theta_{[\nu\rho]}, \quad 
  \partial_\mu J^\mu = 0, 
  \label{eq:eom}
\end{equation}
where $\Theta^{\mu\nu}$, $\Sigma^{\mu}_{~\rho\sigma}$, and $J^\mu $ are the energy-momentum tensor, the spin current, and the charge current, respectively. 
Note that the second equation follows from the total angular momentum conservation $\partial_\mu J^{\mu}_{~\nu\rho} = 0$ and a decomposition of the total angular momentum tensor $J^\mu_{~\nu\rho} = L^\mu_{~\nu\rho} + \Sigma^{\mu}_{~\nu\rho}$ with $L^\mu_{~\nu\rho} = x_\nu \Theta^\mu_{~\rho} - x_\rho \Theta^{\mu}_{~\nu}$ being the orbital angular momentum.  
We choose
the spin current $\Sigma^\mu_{~\nu\rho}$ to be totally anti-symmetric, motivated by Dirac fermions as the microscopic carrier of spin.

We then introduce our dynamical variables --- the energy density $e$, 
the fluid-four velocity $u^\mu$, and the spin density $\sigma_{\nu\rho}$ --- by 
\begin{equation}
 \Theta^{(\mu\nu)} u_\nu = - e u^\mu, \quad
  u_\mu \Sigma^\mu_{~\nu\rho} = - \sigma_{\nu\rho},
  \quad
  J^\mu u_\mu = - n ,
\end{equation}
where we employed the Landau frame in defining the fluid velocity and the energy density~\cite{Landau:Fluid} and normalize the fluid velocity as $u^\mu u_\mu = -1$.  The spin density $\sigma^{\mu\nu}$ only has three dynamical degrees of freedom (rather than six), as the totally anti-symmetric property of the spin current forces $\sigma_{\mu\nu}$ to satisfy the Frenkel condition 
\cite{Frenkel:1926zz} $\sigma_{\mu\nu} u^\nu = 0$ in addition to the anti-symmetric condition $\sigma_{\nu\rho} = - \sigma_{\rho\nu}$.  Consequently, three components of the spin non-conservation law, 
\begin{equation}
 u^\rho \partial_\mu \Sigma^{\mu}_{~\nu\rho} = 
  - 2 \Theta_{[\nu\rho]} u^\rho ,
  \label{eq:constraint-0}
\end{equation}
give constraint equations rather than dynamical ones.
This becomes manifest in the fluid rest frame 
specified as $u^\mu = (1,\bzero)$, 
where Eq.~\eqref{eq:constraint-0} does not have any time derivative.
In the following, for notational simplicity we also use the dual variable $\sigma^\mu \equiv
  \frac{1}{2} \epsilon^{\mu\nu\rho\sigma} u_\nu \sigma_{\rho\sigma} $ 
(or $\sigma_{\mu\nu} = - \epsilon_{\mu\nu\rho\sigma} u^\rho \sigma^\sigma$), 
which satisfies $\sigma^\mu u_\mu = 0$.

We introduce entropy density $s$ as a function of our dynamical variables $e,\sigma_{\mu\nu},n$, and define the conjugate variables --- the inverse temperature $\beta = 1/T$, chemical potential $\mu$, and spin potential $\mu_{\nu\rho}$ --- as
\begin{equation}
 T \diff s 
  = \diff e - \frac{1}{2} \mu^{\nu\rho} \diff \sigma_{\nu\rho} - \mu \diff n 
  \eqq
  \beta \equiv \frac{\partial s}{\partial e}, 
  \quad 
  \beta \mu^{\nu\rho} \equiv - 2 \frac{\partial s}{\partial \sigma_{\nu\rho}}, 
  \quad 
  \beta \mu \equiv - \frac{\partial s}{\partial n}.
\end{equation}
Since the spin density $\sigma_{\nu\rho}$ is transverse to the fluid velocity $u^\mu$, it is sufficient to assume that the spin potential
$\mu^{\nu\rho} (= - \mu^{\rho\nu})$ and its dual $\mu_\mu \equiv \frac{1}{2} \epsilon_{\mu\nu\rho\sigma} u^\nu \mu^{\rho\sigma}$ satisfy $\mu_\mu u^\mu = 0 = \mu^{\mu\nu} u_\nu$.  We will also use the thermal vector $\beta^\mu \equiv \beta u^\mu$ in the subsequent calculations for brevity.

To find the constitutive relations of strongly rotating spinful fluids, 
we rely on a power counting scheme different from that for the conventional hydrodynamics.  
We count vorticity, the other gradients, and spin density 
all differently: 
Spatial components of the (thermal) vorticity 
$\omega_{\mu\nu} \equiv \partial_{[\mu} \beta_{\nu]}$ (or its dual vector $\omega^\mu \equiv \frac{1}{2} \epsilon^{\mu\nu\rho\sigma} u_\nu \omega_{\rho\sigma}$) and the spin potential $\mu^{\nu\rho}$ are counted as $O (\partial^0)$, while their difference ($\omega_{\mu \nu} - \beta \mu_{\mu\nu}$) and the other gradients are counted 
as $O(\partial^1)$. Besides, we assume that the spin density
is suppressed by another small parameter $\eps$ as $\sigma_{\nu\rho} = O (\eps)$, while the vorticity-induced part of the spin potential is not.
This designed power counting scheme is motivated by two observations: (i) vorticity belongs to a family of non-dissipative gradients that does not produce entropy, and (ii) the spin operator is accompanied by the Planck constant since spin is a quantum object. 
As the spin potential is not completely fixed by the vorticity, the three spin densities enjoy their intrinsic dynamics. 

In our power counting scheme, the zeroth-order quantities
that one can use for the tensor decomposition are $u^\mu$, $\home^{\mu} = \omega^\mu/\sqrt{\omega^\nu \omega_\nu}$, 
$\epsilon^{\mu\nu} = \epsilon^{\mu\nu\rho\lambda} u_\rho \home_\lambda$,
and $\eta^{\mu\nu}$.
Assuming the parity symmetry\footnote{For example, the parity-odd term $u^\mu\hat\omega^\nu$ in $\Theta^{\mu\nu}$ is excluded in parity-even systems since the fluid velocity $u^\mu$ and the vorticity $\hat\omega^\mu$ are parity-odd and parity-even, respectively.} and noticing the totally anti-symmetric property of the spin current, one may parametrize the (non-)conserved currents as 
\begin{equation}
 \begin{split}
  \Theta^{\mu\nu}
  &= (e + p_1 ) u^\mu u^\nu + p_1 \eta^{\mu\nu} + p_2 \home^\mu \home^\nu 
  + p_3 \epsilon^{\mu\nu}
  + u^\mu \delta q^\nu - \delta q^\mu u^\nu
  + \delta \Theta^{\mu\nu},
  \\
  \Sigma^{\mu\nu\rho}
  &= \epsilon^{\mu\nu\rho\lambda} 
  (\sigma_\lambda + u_\lambda \delta \sigma),
  \\
  J^\mu 
  &= n u^\mu + \delta J^\mu,
 \end{split}
 \label{eq:parametrized-current}
\end{equation}
where $\delta \Theta^{\mu\nu}$, $\delta q^\mu$, and 
$\delta J^\mu$ satisfy 
$\delta \Theta^{\mu\nu} u_\nu = 0 = u_\mu \delta \Theta^{\mu\nu}$ 
and $\delta q^\mu u_\mu = 0 = \delta J^\mu u_\mu$. 
Note that the constraint equation \eqref{eq:constraint-0} 
with the above parametrization relates $\delta q^\mu$ with the spin density $\sigma^{\mu\nu}$ as  
\begin{eqnarray}
\delta q^\nu 
  = - \frac{1}{2} \epsilon^{\mu\nu\rho\lambda} 
  u_\rho \partial_\mu 
   (\sigma_\lambda + u_\lambda \delta \sigma) 
   .
   \label{eq:q-constraint}
\end{eqnarray}
In the following, we will express 
the so-far undetermined quantities in Eq.~\eqref{eq:parametrized-current} 
in terms of our conjugate 
variables $\beta^\mu, \mu^{\nu\rho}$, and $\mu$. 
Those quantities are $p_a~ (a=1,2,3)$ at $O(1)$ and $\delta \Theta^{\mu\nu}$, $\delta \sigma$, and $\delta J^\mu$ at $O(\partial)$. 
Then, the equations of motion \eqref{eq:eom} serve as a closed set of partial differential equations.

To obtain the constitutive relations, we use the second law of local thermodynamics: 
There exists an entropy current $s^\mu$ satisfying $\partial_\mu s^\mu \geq 0$ for any configuration of $\beta^\mu$, $\mu^{\nu\rho}$, and $\mu$.
Expressing the entropy current as $s^\mu = s u^\mu + \delta s^\mu$ with the first-order correction $\delta s^\mu$, 
one obtains constraints resulting from the condition $\partial_\mu s^\mu \geq 0$ 
up to the second order in derivatives $\partial$ and spin $\hbar$.  This enables us to relate $p_a~ (a=1,2,3)$ to the thermodynamic quantities and determine the tensor structures of $\delta \Theta^{\mu\nu}$, $\delta \sigma$, and $\delta J^\mu$, and $\delta s^\mu$. 
We first insert Eq.~\eqref{eq:parametrized-current} into 
the equations of motion \eqref{eq:eom}. 
Contracting them with $\beta_\nu$, $\frac{1}{2} \beta \mu^{\nu\rho}$, and $\beta \mu$, we have 
\begin{align}
 \beta D e 
 &= - \beta (e + p_1) \theta
 - p_2 \home^\mu \home_\nu \partial_\mu \beta^\nu 
 - p_3 \epsilon^{\mu\nu} \partial_\mu \beta_\nu
 - \delta \Theta^{(\mu\nu)} \partial_{\mu} \beta_{\nu}
 - \delta \Theta^{[\mu\nu]} \partial_{\mu} \beta_{\nu}
 \nonumber \\
 &\quad
 - \delta q_\nu ( D \beta^\nu + \partial^\nu \beta ) + \partial_\mu  (\beta \delta q^\mu ) ,
 \nonumber \\
 - \frac{1}{2} \beta \mu^{\nu\rho} D \sigma_{\nu\rho}
 &= \mu^\mu \sigma_\nu \partial_\mu \beta^\nu  
 - \frac{1}{2} \delta \sigma \epsilon^{\mu\nu\rho\lambda}
 u_\lambda \partial_\mu (\beta \mu_{\nu\rho})
 + \partial_\mu (\beta \mu^\mu \delta \sigma )
  + \Big[p_3 \beta \mu_{\nu\rho} \epsilon^{\nu\rho}
 + \beta \mu_{\mu\nu} \delta \Theta^{[\mu\nu]} \Big]
 ,
 \nonumber \\
 - \beta \mu D n &= \beta \mu n \theta 
 - \delta J^\mu \partial_\mu (\beta \mu)
 + \partial_\mu (\beta \mu \delta J^\mu),
 \label{eq:contracted-eom}
\end{align}
where we introduced $D \equiv u^\mu \partial_\mu$ and $\theta \equiv \partial_\mu u^\mu $ 
and eliminated $\delta q^\mu$ using the constraint equation (\ref{eq:q-constraint}). 
Applying the chain rule for $s (x) = s \big(e (x), \sigma_{\nu\rho} (x),n (x) )$ and using Eq.~\eqref{eq:contracted-eom}, we rewrite the entropy production rate as
\begin{align}
 \partial_\mu s^\mu
 &= s \theta
 + \beta D e
 - \frac{1}{2} \beta \mu^{\nu\rho} D \sigma_{\nu\rho}
 - \beta \mu D n
 + \partial_\mu \delta s^\mu
 \nonumber \\
 &= \left[ s - \beta \left( e + p_1 - \mu n \right) \right] 
 \theta
  - (p_2 \home^{\mu} \home^{\nu} - \mu^\mu \sigma^\nu) \partial_\mu \beta_\nu
 - p_3 \epsilon^{\mu\nu} ( \partial_\mu \beta_\nu - \beta \mu_{\mu\nu} )
 \nonumber \\
 &\quad
 - \delta \Theta^{(\mu\nu)} \partial_{(\mu} \beta_{\nu)}
 - \delta \Theta^{[\mu\nu]} 
 ( \partial_{[\mu} \beta_{\nu]} - \beta \mu_{\mu\nu} )
  - \delta \sigma \partial_{\perp\mu} (\beta \mu^\mu - \omega^\mu) 
  - \delta J^\mu \partial_\mu (\beta \mu)
 \nonumber \\
 &\quad 
 + \partial_\mu 
 \left( 
 \delta s^\mu 
 + \beta \delta q^\mu
 + \beta \mu^\mu \delta \sigma 
 + \beta \mu \delta J^\mu
  \right)
   + \frac{1}{2} \epsilon^{\mu\nu\rho\sigma} u_\rho 
 \partial_\mu \sigma_\sigma (\partial_\nu \beta + D \beta_\nu) ,
 \label{eq:entropy-production-exact}
\end{align}
where we used $\omega^\mu D u_\mu = \partial_{\perp \mu} \omega^\mu - T \omega^\mu \partial_\mu \beta $, following from a Bianchi-like identity
$\partial_\mu (\epsilon^{\mu\nu\rho\sigma} \omega_{\rho\sigma}) = 0$,
and introduced
$\partial_{\perp\mu} \equiv (\delta_\mu^{~\nu} + u_\mu u^\nu) \partial_\nu$.  We also retained all the terms here, 
i.e., Eq.~(\ref{eq:entropy-production-exact}) is exact and does not depend on the power counting scheme at this stage.

To implement our power counting scheme in a clear manner, 
we decompose $\mu^\mu$ and $\sigma_\nu$ 
into the zeroth-order components along $\home^\mu$ and the derivative corrections as
\begin{equation}
 \mu^\mu = \mu_{\para} \home^\mu + \mu_\perp^\mu , \quad 
  \sigma^\nu = \sigma_{\para} \home^\nu + \sigma_{\perp}^\nu,
\end{equation}
where $\sigma_{\para} = O (\eps^1) $ and $\sigma_{\perp}^\mu = O (\eps^1 \partial^1) $ 
according to our power counting scheme. 
On the other hand, the magnitude of the spin potential reads $\mu_{\para} = O (1) $ and $\mu_{\perp}^\mu = O ( \partial^1) $ without the suppression by $\eps$ because the spin potential induced by vorticity is not suppressed according to our assumption.  
The perpendicular components are higher order in derivative because they vanish in the local equilibrium where the entropy production terminates with a condition $\omega_{\mu\nu} = \beta \mu_{\mu\nu}$.
We then obtain the divergence of the entropy current up to $O (\eps^1 \partial^2, \partial^3)$ as 
\begin{align}
 \partial_\mu s^\mu
 &= 
 \left[ s - \beta \left( e + p_1 - \mu n \right) \right] 
 \theta
  - ( p_2 - \mu_{\para} \sigma_{\para} ) \home^\mu \home^\nu
 \partial_\mu \beta_\nu
 - p_3 \epsilon^{\mu\nu} ( \partial_\mu \beta_\nu - \beta \mu_{\mu\nu} )
 \nonumber \\
 &\quad 
 - \delta \Theta^{(\mu\nu)} \partial_{(\mu} \beta_{\nu)}
 - \delta \Theta^{[\mu\nu]} ( \partial_{[\mu} \beta_{\nu]} - \beta \mu_{\mu\nu} )
 - \delta J^\mu \partial_\mu (\beta \mu)
 \nonumber \\
 &\quad 
 + \partial_\mu 
 \left( 
 \delta s^\mu 
 + \beta \mu^\mu \delta \sigma 
 + \beta \mu \delta J^\mu
  \right)
  + O (\eps^1 \partial^2, \partial^3) .
 \label{eq:entropy-production-approximated}
\end{align}
Note that here one can safely drop the term $\delta \sigma \partial_{\perp\mu} (\beta \mu^\mu - \omega^\mu)$ in Eq.~(\ref{eq:entropy-production-exact}), which is higher-order in derivatives.  

\sect{Constitutive relations of rotating spinful fluids}
We derive the constitutive relations up to the first order in derivatives $\partial$ and spin $\hbar$ by requiring that the divergence of the entropy current~\eqref{eq:entropy-production-approximated} respects the second law of thermodynamics.

Let us first consider the leading-order constitutive relation for the energy-momentum tensor.
The first line of Eq.~\eqref{eq:entropy-production-approximated} results from the non-dissipative terms, which we require to vanish so that $\partial_\mu s^\mu = 0$ at the leading order.  We find
\begin{equation}
 e + p_1 = Ts + \mu n , \quad
  p_2 = \mu_{\para} \sigma_{\para}, \quad 
  p_3 = 0 ,
  \label{eq:Constraints-1}
\end{equation}
which relates the parameters $p_a~(a=1,2,3)$ with the thermodynamic quantities.  Equation~(\ref{eq:Constraints-1}) indicates that the leading-order energy-momentum tensor has
anistropic pressure induced by vorticity as
\begin{equation}
\label{eqideal}
 \Theta^{\mu\nu}_{(0)}
  = e u^\mu u^\nu + p_{\perp} \Xi^{\mu\nu} + p_{\para} \home^\mu \home^\nu, 
\end{equation}
where we introduced $p_{\perp} \equiv p_1$ and  $p_{\para} \equiv p_1 + \mu_{\para} \sigma_{\para}$ and the projection tensor $\Xi^{\mu\nu} \equiv \eta^{\mu\nu} + u^\mu u^\nu - \home^\mu \home^\nu$.
Note that one can express the first relation of Eq.~\eqref{eq:Constraints-1} as
$e+ p_{\para} = Ts + \mu n + \mu_{\para} \sigma_{\para}$ that captures the contribution of spin to the thermodynamic relation.  The constitutive relation (\ref{eqideal}) bears similarity with that for relativistic magnetohydrodynamics in which the direction of magnetic field plays the role of $\home^\mu$ and magnetization of the fluid induces difference between $p_\perp$ and $p_\para$~\cite{Huang:2011dc,Finazzo:2016mhm,Grozdanov:2016tdf,Hernandez:2017mch,Hongo:2020qpv}.

We turn to the the first-order corrections to the constitutive relations.  Requiring the local second law of thermodynamics $\partial_\mu s^\mu \geq 0$, we find
\begin{equation}
 \begin{split}
  \delta \Theta^{(\mu\nu)}
  &= - T  \eta^{\mu\nu\rho\sigma}   \partial_{(\rho} \beta_{\sigma)}
   - T   \xi^{\mu\nu\rho\sigma}  
  ( \partial_{[\rho} \beta_{\sigma]} - \beta \mu_{\rho\sigma} ) 
  ,
  \\
  \delta \Theta^{[\mu\nu]}
  &= - T   \gamma^{\mu\nu\rho\sigma}    ( \partial_{[\rho} \beta_{\sigma]} - \beta \mu_{\rho\sigma} ) 
  - T \xi^{\prime \, \mu\nu\rho\sigma}  \partial_{(\rho} \beta_{\sigma)}    ,
  \\
  \delta J^\mu
  &= - T \kappa^{\mu\nu} \partial_\nu (\beta \mu),
  \\
  \delta s^\mu
  &= 
  - \beta \mu^\mu \delta \sigma 
  - \beta \mu \delta J^\mu,
 \end{split}
\end{equation}
with the two viscous tensors $\eta^{\mu\nu\rho\sigma}$ and 
$\gamma^{\mu\nu\rho\sigma}$, and conductivity tensor $\kappa^{\mu\nu}$. 
In addition, there are ``cross terms'' given by $ \xi^{\mu\nu\rho\sigma}  $ 
and $ \xi^{\prime \, \mu\nu\rho\sigma}  $ 
that convert the symmetric and antisymmetric flow gradients 
to the antisymmetric and symmetric parts of $   \delta \Theta^{\mu\nu} $, respectively.\footnote{
The presence of the cross terms is expected from 
the properties of the Coriolis force $ {\bm F} \propto {\bm v} \times {\bm \omega} $ 
with the velocity $ {\bm v}  $ and angular velocity $ {\bm \omega}   $. 
Operating derivatives on the Coriolis force, one finds a nonzero divergence and rotation of the force, 
$  \nabla \cdot {\bm F} \propto   {\bm \omega}  \cdot (\nabla \times {\bm v} )$ and $
\nabla \times {\bm F} \propto  ( {\bm \omega}  \cdot \nabla ) {\bm v} -  {\bm \omega}  (\nabla \cdot {\bm v} )  $, 
induced by the rotation and gradient/divergence of $ {\bm v} $, respectively. 
Here, we assumed a constant $ {\bm \omega} $ for demonstration. } 
Similar cross terms have been discussed in the context of nematic liquid crystal~\cite{Parodi1970StressTF} (see also Ref.~\cite{RevModPhys.46.617} for a review). 

These tensors are decomposed by the use of the order-one spatial tensors $\home^\mu$, $\Xi^{\mu\nu}$, and $\epsilon^{\mu\nu}$. 
The transverse, longitudinal, and Hall-like components read 
\begin{equation}
 \begin{split}
  \eta^{\mu\nu\rho\sigma}
  &= \zeta_{\perp} \Xi^{\mu\nu} \Xi^{\rho\sigma}
  + \zeta_{\para} \home^\mu \home^\nu \home^{\rho} \home^\sigma
  + \zeta_{\times} 
  \big( \home^\mu \home^\nu \Xi^{\rho\sigma} + \Xi^{\mu\nu} \home^\rho \home^\sigma
  \big)
  \\
  &\quad 
  + \eta_{\perp}
  \big( 
  \Xi^{\mu\rho} \Xi^{\nu\sigma} 
  + \Xi^{\mu\sigma} \Xi^{\nu\rho} 
  - \Xi^{\mu\nu} \Xi^{\rho\sigma}  
  \big)
  + 2 \eta_{\para} 
  \big( \home^\mu \Xi^{\nu(\rho} \home^{\sigma)}
  + \home^\nu \Xi^{\mu(\rho} \home^{\sigma)} 
  \big)
  \\
  &\quad 
  + 2 \eta_{H_\perp} 
  \big( 
  \Xi^{\mu(\rho} \epsilon^{\sigma)\nu}
  + \Xi^{\nu(\rho} \epsilon^{\sigma)\mu}
  \big)
  + 2 \eta_{H_\para} 
  \big( 
  \home^\mu \epsilon^{\nu(\rho} \home^{\sigma)}
  + \home^\nu \epsilon^{\mu(\rho} \home^{\sigma)}
  \big),
  \\
  \gamma^{\mu\nu\rho\sigma}
  &=
  \gamma_\perp 
  \big(
  \Xi^{\mu\rho} \Xi^{\nu\sigma} - \Xi^{\mu\sigma} \Xi^{\nu\rho}
  \big)
  + 2 \gamma_{\para}   
  \big( 
  \home^\mu \Xi^{\nu[\rho} \home^{\sigma]}
  -  \home^\nu \Xi^{\mu[\rho} \home^{\sigma]} 
  \big)
  \\
  &\quad 
  + 2  \gamma_{H } 
  \big( 
  \home^\mu \epsilon^{\nu[\rho} \home^{\sigma]}
  - \home^\nu \epsilon^{\mu[\rho} \home^{\sigma]}
  \big),
  \\
 \xi^{\mu\nu\rho\sigma} 
&=
  2 \xi_\para \big( \home^\mu \Xi^{\nu [\rho} \home^{\sigma]} + \home^\nu \Xi^{\mu [\rho} \home^{\sigma]}
+  \home^\mu \Xi^{\nu (\rho} \home^{\sigma)} - \home^\nu \Xi^{\mu (\rho} \home^{\sigma)} \big)
   \\
 &\quad
  +  \zeta_{H_\perp}   
  \big(  \Xi^{\mu\nu}  \epsilon^{\rho\sigma} -  \Xi^{\rho \sigma}  \epsilon^{\mu\nu} \big)
  +  \zeta_{H\para} \big(  \home^\mu \home^{\nu}  \epsilon^{\rho\sigma} -  \home^\rho \home^{\sigma}  \epsilon^{\mu\nu}  \big)
  \\
 &\quad
  + 2 \xi_{H}   \big(   \home^\mu \epsilon^{\nu[\rho} \home^{\sigma]}  + \home^\nu \epsilon^{\mu[\rho} \home^{\sigma]} 
+  \home^\mu \epsilon^{\nu(\rho} \home^{\sigma)}  - \home^\nu \epsilon^{\mu(\rho} \home^{\sigma)}  \big)
,
\\
  \kappa^{\mu\nu}
  &= \kappa_{\perp} \Xi^{\mu\nu}
  + \kappa_{\para} \home^\mu \home^\nu
  + \kappa_H \epsilon^{\mu\nu}.
 \end{split}
\end{equation}
Those viscous tensor should satisfy the Onsager reciprocal relation 
stemming from the time-reversal symmetry~\cite{Onsager1931, de1984non}. 
Note that $ \home $ is an odd quantity under the time-reversal transformation. 
Then, the above tensors should satisfy the Onsager reciprocal relations 
$   \eta^{\mu\nu\rho\sigma} (\home) =   \eta^{\rho\sigma\mu\nu} (-\home) $, 
$   \gamma^{\mu\nu\rho\sigma} (\home) =   \gamma^{\rho\sigma\mu\nu} (-\home) $, 
$   \xi^{\mu\nu\rho\sigma} (\home) =   \xi^{\rho\sigma\mu\nu} (-\home) $, 
and $\xi^{\prime \, \mu\nu\rho\sigma} (\home) =  \xi^{\rho\sigma\mu\nu} (- \home)  $. 
The last relation provides $ \xi^{\prime \, \mu\nu\rho\sigma}  $ in terms of $ \xi^{\mu\nu\rho\sigma}  $. 

The terms appearing in the viscous and conductivity tensors can be classified into two types according to their properties under the interchange between (pairs of) Lorentz indices --- $ (\mu,\nu)\leftrightarrow(\rho,\sigma)$ for the viscous tensors and $\mu \leftrightarrow \nu$ for the conductivity tensor --- without flipping the sign of $ \home $.
The first type, the terms having the subscript $ H $, is anti-symmetric under the index interchange.
These are non-dissipative Hall-like terms, which do not contribute to the entropy production. The other terms, i.e., those without the subscript $H$,
are symmetric under the index interchange, 
meaning that they are dissipative and contribute to the entropy production. 
Among these, the bulk viscosity $\zeta_\times$ captures the non-equilibrium correction to the anisotropic pressures $p_\perp$ and $p_\parallel$ in response to expansion or compression in the parallel and perpendicular directions, respectively. 
Those two processes are reciprocal to one another, 
and we required them to satisfy the Onsager reciprocal relation. 
As a result, we find seventeen transport coefficients:
three bulk viscosities $\zeta_{\para}, \zeta_{\perp}, \zeta_{\times}$, 
four shear viscosities $\eta_{\para}, \eta_{\perp}, \eta_{H_{\para}}, \eta_{H_{\perp}}$, three rotational viscosities $\gamma_{\para}, \gamma_{\perp}, \gamma_{H}$, 
four cross viscosities $ \xi_\para, \zeta_{H\para}, \zeta_{H\perp}, \xi_H $ and three conductivities $\kappa_{\perp}, \kappa_{\para}, \kappa_H$. 
Note that we do not include a possible transverse structure, $ \Xi^{\mu[\rho} \epsilon^{\sigma]\nu}   - \Xi^{\nu[\rho} \epsilon^{\sigma]\mu}$, in $\gamma^{\mu\nu\rho\sigma}$ since it identically vanishes according to the so-called Schouten identity $\eta^{\mu\gamma} \epsilon^{\rho\nu\alpha\beta} - \eta^{\rho \gamma} \epsilon^{\mu\nu\alpha\beta} - \eta^{\nu \gamma} \epsilon^{\rho\mu\alpha\beta} - \eta^{\alpha\gamma} \epsilon^{\rho\nu\mu\beta}
-\eta^{\beta\gamma} \epsilon^{\rho\nu\alpha\mu} =0 $.

To respect the second law, the dissipative transport coefficients must satisfy a set of inequalities,
\begin{equation}
 \begin{split}
  &\zeta_{\perp} \geq 0 , \quad
  \zeta_{\para} \geq 0, \quad 
  \zeta_{\perp} \zeta_{\para} - \zeta_{\times}^2 \geq 0 , \quad 
  \\
  &\eta_{\perp} \geq 0, \quad 
  \eta_{\para} \geq 0 , \quad 
  \gamma_{\perp} \geq 0 , \quad 
  \gamma_{\para} \geq 0 , \quad
  \eta_{\para} \gamma_{\para} - \xi_{\para}^2 \geq 0  , \quad
  \\ 
  & \kappa_{\perp} \geq 0, \quad 
  \kappa_{\para} \geq 0 . 
 \end{split}
 \label{eq:positivity-constraint}
\end{equation} 
On the other hand, the signs of the seven coefficients $\eta_{H_\perp}, \eta_{H_\para}, \gamma_{H},
 \zeta_{H\para}, \zeta_{H\perp}, \xi_H,$
and $\kappa_H$ are not constrained by the second law,
as they do not contribute to the entropy production just like the Hall transports in relativistic magnetohydrodynamics~\cite{Huang:2011dc,Finazzo:2016mhm,Hernandez:2017mch,Grozdanov:2016tdf,Hongo:2020qpv}. 
Those non-dissipative transports arise due to strong vorticity, which can be interpreted as an analogue of strong magnetic fields~\cite{Landau:Kinetics}: 
The Coriolis force does not do work, similarly to the Lorentz force by magnetic fields. 
One clear difference from the Hall transports in magnetohydrodynamics is that those transport coefficients in gyrohydrodynamics are even under charge conjugation 
and can have non-vanishing values for neutral fluids.

\sect{Summary and Outlook}
In this Letter, we have reported the phenomenological derivation of the constitutive relations for spinful fluids with strong vorticity. 
It is formulated in the ``spin hydrodynamic regime,'' where spin density enjoys its intrinsic dynamics as the energy-momentum and vector charge densities do.   
We have found the leading and next-to-leading order constitutive relations that capture the spin dynamics intertwined with rotating fluid dynamics as well as the anisotropic pressure and transport phenomena induced by the strong vorticity.
We have obtained seventeen transport coefficients: three bulk viscosities $\zeta_{\para}, \zeta_{\perp}, \zeta_{\times}$, 
four shear viscosities $\eta_{\para}, \eta_{\perp}, \eta_{H_{\para}}, \eta_{H_{\perp}}$, three rotational viscosities $\gamma_{\para}, \gamma_{\perp}, \gamma_{H}$, 
four cross viscosities $ \xi_\para, \zeta_{H\para}, \zeta_{H\perp}, \xi_H $ and three conductivities $\kappa_{\perp}, \kappa_{\para}, \kappa_H$.  A subset of the transport coefficients should satisfy the positivity constraint~\eqref{eq:positivity-constraint}, while there are no such constraints on the Hall-like transport coefficients.

The entropy-current analysis leaves the seventeen transport coefficients as phenomenological parameters.
The theoretical evaluation of these parameters 
can be carried out on the basis of either 
the linear-mode analysis or the Green-Kubo formula~\cite{Green,Nakano,Kubo,Kadanoff-Martin1963,Luttinger:1964zz}, 
both of which are regarded as the matching conditions for the (quasi-)hydrodynamic description. 
Deriving such matching conditions for the seventeen transport coefficients and evaluating them with  microscopic theories (such as quantum chromodynamics) are important outlooks in line with the present Letter.

We investigated the formulation with the totally anti-symmetric spin current that is motivated by the canonical spin current of Dirac fermions,
while previous works do not necessarily assume this tensor structure (e.g., Ref.~\cite{Hattori:2019lfp}).
One may always redefine the spin current and energy-momentum tensor operators --- or change what one calls spin current and energy-momentum tensor operators --- without modifying the form of their equations of motion via the so-called pseudo-gauge transformations~\cite{Belinfante1939,Belinfante1940,Rosenfeld1940} 
(see, e.g., Refs.~\cite{Becattini:2018duy, Li:2020eon, Speranza:2020ilk} for recent discussion).
This may lead to another definition of a spin current operator which is not necessarily totally anti-symmetric.
It may be worth investigating what the resulting hydrodynamic equation looks like with a different choice of the spin and energy-momentum tensor.
Besides, strong vorticity can give rise to intriguing phenomena in spinless case as well. Interested readers can find discussions in Ref. \cite{Li:2020eon} for possible anisotropies in constitutive relations and in, e.g., Ref.~\cite{Jensen:2011xb} for the modification of thermodynamics in a $2+1$ dimensional fluid. We leave all these issues as future works.

\section*{Acknowledgment}
The authors thank Matthias Kaminski, Yu-Chen~Liu, Kazuya~Mameda, Misha~Stephanov, Xiao-Liang~Xia and Ho-Ung~Yee for fruitful discussions.
This work is partially supported by JSPS KAKENHI under grant Nos.~20K03948, 22H02316, and 22K14045 and RIKEN iTHEMS Program (in particular iTHEMS Non-Equilibrium Working group and iTHEMS Mathematical Physics Working group). Z.~C. and X.-G.~H. are supported by NSFC under Grant No.~12075061 and Shanghai NSF under Grant No.~20ZR1404100.

\bibliographystyle{utphys}
\bibliography{refs}

\providecommand{\href}[2]{#2}\begingroup\raggedright\begin{thebibliography}{10}

\bibitem{Landau:Fluid}
L.~D. Landau and E.~M. Lifshitz, {\em {Fluid Mechanics, Second Edition}}.
\newblock Butterworth Heinemann, Oxford, UK, 1987.

\bibitem{Camenzind:book}
M.~Camenzind, {\em {Compact Objects in Astrophysics}}.
\newblock Springer, 2007.

\bibitem{Rezzolla:book}
L.~Rezzolla and O.~Zanotti, {\em {Relativistic Hydrodynamics}}.
\newblock Oxford University Press, 2013.

\bibitem{Baiotti:2016qnr}
L.~Baiotti and L.~Rezzolla, ``{Binary neutron star mergers: a review of
  Einstein\textquoteright{}s richest laboratory},''
  \href{http://dx.doi.org/10.1088/1361-6633/aa67bb}{{\em Rept. Prog. Phys.}
  {\bfseries 80} no.~9, (2017) 096901},
  \href{http://arxiv.org/abs/1607.03540}{{\ttfamily arXiv:1607.03540 [gr-qc]}}.

\bibitem{Paschalidis:2016vmz}
V.~Paschalidis and N.~Stergioulas, ``{Rotating Stars in Relativity},''
  \href{http://dx.doi.org/10.1007/s41114-017-0008-x}{{\em Living Rev. Rel.}
  {\bfseries 20} no.~1, (2017) 7},
  \href{http://arxiv.org/abs/1612.03050}{{\ttfamily arXiv:1612.03050
  [astro-ph.HE]}}.

\bibitem{Muller:2020ard}
B.~M\"uller, ``{Hydrodynamics of core-collapse supernovae and their
  progenitors},'' \href{http://dx.doi.org/10.1007/s41115-020-0008-5}{{\em
  Astrophysics} {\bfseries 6} (2020) 3},
  \href{http://arxiv.org/abs/2006.05083}{{\ttfamily arXiv:2006.05083
  [astro-ph.SR]}}.

\bibitem{Heinz:2013th}
U.~Heinz and R.~Snellings, ``{Collective flow and viscosity in relativistic
  heavy-ion collisions},''
  \href{http://dx.doi.org/10.1146/annurev-nucl-102212-170540}{{\em Ann. Rev.
  Nucl. Part. Sci.} {\bfseries 63} (2013) 123--151},
  \href{http://arxiv.org/abs/1301.2826}{{\ttfamily arXiv:1301.2826 [nucl-th]}}.

\bibitem{Florkowski:2017olj}
W.~Florkowski, M.~P. Heller, and M.~Spalinski, ``{New theories of relativistic
  hydrodynamics in the LHC era},''
  \href{http://dx.doi.org/10.1088/1361-6633/aaa091}{{\em Rept. Prog. Phys.}
  {\bfseries 81} no.~4, (2018) 046001},
  \href{http://arxiv.org/abs/1707.02282}{{\ttfamily arXiv:1707.02282
  [hep-ph]}}.

\bibitem{Shen:2020mgh}
C.~Shen and L.~Yan, ``{Recent development of hydrodynamic modeling in heavy-ion
  collisions},'' \href{http://dx.doi.org/10.1007/s41365-020-00829-z}{{\em Nucl.
  Sci. Tech.} {\bfseries 31} no.~12, (10, 2020) 122},
  \href{http://arxiv.org/abs/2010.12377}{{\ttfamily arXiv:2010.12377
  [nucl-th]}}.

\bibitem{STAR:2017ckg}
{\bfseries STAR} Collaboration, L.~Adamczyk {\em et~al.}, ``{Global $\Lambda$
  hyperon polarization in nuclear collisions: evidence for the most vortical
  fluid},'' \href{http://dx.doi.org/10.1038/nature23004}{{\em Nature}
  {\bfseries 548} (2017) 62--65},
  \href{http://arxiv.org/abs/1701.06657}{{\ttfamily arXiv:1701.06657
  [nucl-ex]}}.

\bibitem{Adam:2018ivw}
{\bfseries STAR} Collaboration, J.~Adam {\em et~al.}, ``{Global polarization of
  $\Lambda$ hyperons in Au+Au collisions at $\sqrt{s_{_{NN}}}$ = 200 GeV},''
  \href{http://dx.doi.org/10.1103/PhysRevC.98.014910}{{\em Phys. Rev. C}
  {\bfseries 98} (2018) 014910},
  \href{http://arxiv.org/abs/1805.04400}{{\ttfamily arXiv:1805.04400
  [nucl-ex]}}.

\bibitem{Adam:2020pti}
{\bfseries STAR} Collaboration, J.~Adam {\em et~al.}, ``{Global Polarization of
  $\Xi$ and $\Omega$ Hyperons in Au+Au Collisions at $\sqrt {s_{NN}}$ = 200
  GeV},'' \href{http://dx.doi.org/10.1103/PhysRevLett.126.162301}{{\em Phys.
  Rev. Lett.} {\bfseries 126} no.~16, (2021) 162301},
  \href{http://arxiv.org/abs/2012.13601}{{\ttfamily arXiv:2012.13601
  [nucl-ex]}}.

\bibitem{Acharya:2019vpe}
{\bfseries ALICE} Collaboration, S.~Acharya {\em et~al.}, ``{Evidence of
  Spin-Orbital Angular Momentum Interactions in Relativistic Heavy-Ion
  Collisions},'' \href{http://dx.doi.org/10.1103/PhysRevLett.125.012301}{{\em
  Phys. Rev. Lett.} {\bfseries 125} no.~1, (2020) 012301},
  \href{http://arxiv.org/abs/1910.14408}{{\ttfamily arXiv:1910.14408
  [nucl-ex]}}.

\bibitem{STAR:2022fan}
{\bfseries STAR} Collaboration, M.~Abdallah {\em et~al.}, ``{Observation of
  Global Spin Alignment of $\phi$ and $K^{*0}$ Vector Mesons in Nuclear
  Collisions},'' \href{http://arxiv.org/abs/2204.02302}{{\ttfamily
  arXiv:2204.02302 [hep-ph]}}.

\bibitem{ALICE:2022sli}
{\bfseries ALICE} Collaboration, ``{Measurement of the J/$\psi$ polarization
  with respect to the event plane in Pb-Pb collisions at the LHC},''
  \href{http://arxiv.org/abs/2204.10171}{{\ttfamily arXiv:2204.10171
  [nucl-ex]}}.

\bibitem{Liang:2004ph}
Z.-T. Liang and X.-N. Wang, ``{Globally polarized quark-gluon plasma in
  non-central A+A collisions},''
  \href{http://dx.doi.org/10.1103/PhysRevLett.94.102301}{{\em Phys. Rev. Lett.}
  {\bfseries 94} (2005) 102301},
  \href{http://arxiv.org/abs/nucl-th/0410079}{{\ttfamily
  arXiv:nucl-th/0410079}}. [Erratum: Phys.Rev.Lett. 96, 039901 (2006)].

\bibitem{Liang:2004xn}
Z.-T. Liang and X.-N. Wang, ``{Spin alignment of vector mesons in non-central
  A+A collisions},''
  \href{http://dx.doi.org/10.1016/j.physletb.2005.09.060}{{\em Phys. Lett. B}
  {\bfseries 629} (2005) 20--26},
  \href{http://arxiv.org/abs/nucl-th/0411101}{{\ttfamily
  arXiv:nucl-th/0411101}}.

\bibitem{Betz:2007kg}
B.~Betz, M.~Gyulassy, and G.~Torrieri, ``{Polarization probes of vorticity in
  heavy ion collisions},''
  \href{http://dx.doi.org/10.1103/PhysRevC.76.044901}{{\em Phys. Rev. C}
  {\bfseries 76} (2007) 044901},
  \href{http://arxiv.org/abs/0708.0035}{{\ttfamily arXiv:0708.0035 [nucl-th]}}.

\bibitem{Becattini:2007sr}
F.~Becattini, F.~Piccinini, and J.~Rizzo, ``{Angular momentum conservation in
  heavy ion collisions at very high energy},''
  \href{http://dx.doi.org/10.1103/PhysRevC.77.024906}{{\em Phys. Rev. C}
  {\bfseries 77} (2008) 024906},
  \href{http://arxiv.org/abs/0711.1253}{{\ttfamily arXiv:0711.1253 [nucl-th]}}.

\bibitem{Huang:2011ru}
X.-G. Huang, P.~Huovinen, and X.-N. Wang, ``{Quark Polarization in a Viscous
  Quark-Gluon Plasma},''
  \href{http://dx.doi.org/10.1103/PhysRevC.84.054910}{{\em Phys. Rev. C}
  {\bfseries 84} (2011) 054910},
  \href{http://arxiv.org/abs/1108.5649}{{\ttfamily arXiv:1108.5649 [nucl-th]}}.

\bibitem{Becattini:2013fla}
F.~Becattini, V.~Chandra, L.~Del~Zanna, and E.~Grossi, ``{Relativistic
  distribution function for particles with spin at local thermodynamical
  equilibrium},'' \href{http://dx.doi.org/10.1016/j.aop.2013.07.004}{{\em
  Annals Phys.} {\bfseries 338} (2013) 32--49},
  \href{http://arxiv.org/abs/1303.3431}{{\ttfamily arXiv:1303.3431 [nucl-th]}}.

\bibitem{Liu:2020ymh}
Y.-C. Liu and X.-G. Huang, ``{Anomalous chiral transports and spin polarization
  in heavy-ion collisions},''
  \href{http://dx.doi.org/10.1007/s41365-020-00764-z}{{\em Nucl. Sci. Tech.}
  {\bfseries 31} no.~6, (2020) 56},
  \href{http://arxiv.org/abs/2003.12482}{{\ttfamily arXiv:2003.12482
  [nucl-th]}}.

\bibitem{Gao:2020vbh}
J.-H. Gao, G.-L. Ma, S.~Pu, and Q.~Wang, ``{Recent developments in chiral and
  spin polarization effects in heavy-ion collisions},''
  \href{http://dx.doi.org/10.1007/s41365-020-00801-x}{{\em Nucl. Sci. Tech.}
  {\bfseries 31} no.~9, (2020) 90},
  \href{http://arxiv.org/abs/2005.10432}{{\ttfamily arXiv:2005.10432
  [hep-ph]}}.

\bibitem{Huang:2020xyr}
X.-G. Huang, ``{Vorticity and Spin Polarization \textemdash{} A Theoretical
  Perspective},'' \href{http://dx.doi.org/10.1016/j.nuclphysa.2020.121752}{{\em
  Nucl. Phys. A} {\bfseries 1005} (2021) 121752},
  \href{http://arxiv.org/abs/2002.07549}{{\ttfamily arXiv:2002.07549
  [nucl-th]}}.

\bibitem{Becattini:2020ngo}
F.~Becattini and M.~A. Lisa, ``{Polarization and Vorticity in the
  Quark\textendash{}Gluon Plasma},''
  \href{http://dx.doi.org/10.1146/annurev-nucl-021920-095245}{{\em Ann. Rev.
  Nucl. Part. Sci.} {\bfseries 70} (2020) 395--423},
  \href{http://arxiv.org/abs/2003.03640}{{\ttfamily arXiv:2003.03640
  [nucl-ex]}}.

\bibitem{Becattini:2022zvf}
F.~Becattini, ``{Spin and polarization: a new direction in relativistic heavy
  ion physics},'' \href{http://arxiv.org/abs/2204.01144}{{\ttfamily
  arXiv:2204.01144 [nucl-th]}}.

\bibitem{Montenegro:2017rbu}
D.~Montenegro, L.~Tinti, and G.~Torrieri, ``{Ideal relativistic fluid limit for
  a medium with polarization},''
  \href{http://dx.doi.org/10.1103/PhysRevD.96.056012}{{\em Phys. Rev. D}
  {\bfseries 96} no.~5, (2017) 056012},
  \href{http://arxiv.org/abs/1701.08263}{{\ttfamily arXiv:1701.08263
  [hep-th]}}. [Addendum: Phys.Rev.D 96, 079901 (2017)].

\bibitem{Montenegro:2017lvf}
D.~Montenegro, L.~Tinti, and G.~Torrieri, ``{Sound waves and vortices in a
  polarized relativistic fluid},''
  \href{http://dx.doi.org/10.1103/PhysRevD.96.076016}{{\em Phys. Rev. D}
  {\bfseries 96} no.~7, (2017) 076016},
  \href{http://arxiv.org/abs/1703.03079}{{\ttfamily arXiv:1703.03079
  [hep-th]}}.

\bibitem{Florkowski:2017ruc}
W.~Florkowski, B.~Friman, A.~Jaiswal, and E.~Speranza, ``{Relativistic fluid
  dynamics with spin},''
  \href{http://dx.doi.org/10.1103/PhysRevC.97.041901}{{\em Phys. Rev. C}
  {\bfseries 97} no.~4, (2018) 041901},
  \href{http://arxiv.org/abs/1705.00587}{{\ttfamily arXiv:1705.00587
  [nucl-th]}}.

\bibitem{Florkowski:2018fap}
W.~Florkowski, A.~Kumar, and R.~Ryblewski, ``{Relativistic hydrodynamics for
  spin-polarized fluids},''
  \href{http://dx.doi.org/10.1016/j.ppnp.2019.07.001}{{\em Prog. Part. Nucl.
  Phys.} {\bfseries 108} (2019) 103709},
  \href{http://arxiv.org/abs/1811.04409}{{\ttfamily arXiv:1811.04409
  [nucl-th]}}.

\bibitem{Montenegro:2018bcf}
D.~Montenegro and G.~Torrieri, ``{Causality and dissipation in relativistic
  polarizable fluids},''
  \href{http://dx.doi.org/10.1103/PhysRevD.100.056011}{{\em Phys. Rev. D}
  {\bfseries 100} no.~5, (2019) 056011},
  \href{http://arxiv.org/abs/1807.02796}{{\ttfamily arXiv:1807.02796
  [hep-th]}}.

\bibitem{Hattori:2019lfp}
K.~Hattori, M.~Hongo, X.-G. Huang, M.~Matsuo, and H.~Taya, ``{Fate of spin
  polarization in a relativistic fluid: An entropy-current analysis},''
  \href{http://dx.doi.org/10.1016/j.physletb.2019.05.040}{{\em Phys. Lett. B}
  {\bfseries 795} (2019) 100--106},
  \href{http://arxiv.org/abs/1901.06615}{{\ttfamily arXiv:1901.06615
  [hep-th]}}.

\bibitem{Li:2019qkf}
S.~Li and H.-U. Yee, ``{Quantum Kinetic Theory of Spin Polarization of Massive
  Quarks in Perturbative QCD: Leading Log},''
  \href{http://dx.doi.org/10.1103/PhysRevD.100.056022}{{\em Phys. Rev. D}
  {\bfseries 100} no.~5, (2019) 056022},
  \href{http://arxiv.org/abs/1905.10463}{{\ttfamily arXiv:1905.10463
  [hep-ph]}}.

\bibitem{Fukushima:2020ucl}
K.~Fukushima and S.~Pu, ``{Spin hydrodynamics and symmetric energy-momentum
  tensors \textendash{} A current induced by the spin vorticity
  \textendash{}},''
  \href{http://dx.doi.org/10.1016/j.physletb.2021.136346}{{\em Phys. Lett. B}
  {\bfseries 817} (2021) 136346},
  \href{http://arxiv.org/abs/2010.01608}{{\ttfamily arXiv:2010.01608
  [hep-th]}}.

\bibitem{Bhadury:2020puc}
S.~Bhadury, W.~Florkowski, A.~Jaiswal, A.~Kumar, and R.~Ryblewski,
  ``{Relativistic dissipative spin dynamics in the relaxation time
  approximation},''
  \href{http://dx.doi.org/10.1016/j.physletb.2021.136096}{{\em Phys. Lett. B}
  {\bfseries 814} (2021) 136096},
  \href{http://arxiv.org/abs/2002.03937}{{\ttfamily arXiv:2002.03937
  [hep-ph]}}.

\bibitem{Montenegro:2020paq}
D.~Montenegro and G.~Torrieri, ``{Linear response theory and effective action
  of relativistic hydrodynamics with spin},''
  \href{http://dx.doi.org/10.1103/PhysRevD.102.036007}{{\em Phys. Rev. D}
  {\bfseries 102} no.~3, (2020) 036007},
  \href{http://arxiv.org/abs/2004.10195}{{\ttfamily arXiv:2004.10195
  [hep-th]}}.

\bibitem{Li:2020eon}
S.~Li, M.~A. Stephanov, and H.-U. Yee, ``{Nondissipative Second-Order
  Transport, Spin, and Pseudogauge Transformations in Hydrodynamics},''
  \href{http://dx.doi.org/10.1103/PhysRevLett.127.082302}{{\em Phys. Rev.
  Lett.} {\bfseries 127} no.~8, (2021) 082302},
  \href{http://arxiv.org/abs/2011.12318}{{\ttfamily arXiv:2011.12318
  [hep-th]}}.

\bibitem{Shi:2020htn}
S.~Shi, C.~Gale, and S.~Jeon, ``{From chiral kinetic theory to relativistic
  viscous spin hydrodynamics},''
  \href{http://dx.doi.org/10.1103/PhysRevC.103.044906}{{\em Phys. Rev. C}
  {\bfseries 103} no.~4, (2021) 044906},
  \href{http://arxiv.org/abs/2008.08618}{{\ttfamily arXiv:2008.08618
  [nucl-th]}}.

\bibitem{Garbiso:2020puw}
M.~Garbiso and M.~Kaminski, ``{Hydrodynamics of simply spinning black holes \&
  hydrodynamics for spinning quantum fluids},''
  \href{http://dx.doi.org/10.1007/JHEP12(2020)112}{{\em JHEP} {\bfseries 12}
  (2020) 112}, \href{http://arxiv.org/abs/2007.04345}{{\ttfamily
  arXiv:2007.04345 [hep-th]}}.

\bibitem{Gallegos:2020otk}
A.~D. Gallegos and U.~G\"ursoy, ``{Holographic spin liquids and Lovelock
  Chern-Simons gravity},''
  \href{http://dx.doi.org/10.1007/JHEP11(2020)151}{{\em JHEP} {\bfseries 11}
  (2020) 151}, \href{http://arxiv.org/abs/2004.05148}{{\ttfamily
  arXiv:2004.05148 [hep-th]}}.

\bibitem{She:2021lhe}
D.~She, A.~Huang, D.~Hou, and J.~Liao, ``{Relativistic Viscous Hydrodynamics
  with Angular Momentum},'' \href{http://arxiv.org/abs/2105.04060}{{\ttfamily
  arXiv:2105.04060 [nucl-th]}}.

\bibitem{Hu:2021lnx}
J.~Hu, ``{Kubo formulae for first-order spin hydrodynamics},''
  \href{http://dx.doi.org/10.1103/PhysRevD.103.116015}{{\em Phys. Rev. D}
  {\bfseries 103} no.~11, (2021) 116015},
  \href{http://arxiv.org/abs/2101.08440}{{\ttfamily arXiv:2101.08440
  [hep-ph]}}.

\bibitem{Gallegos:2021bzp}
A.~D. Gallegos, U.~G\"ursoy, and A.~Yarom, ``{Hydrodynamics of spin
  currents},'' \href{http://dx.doi.org/10.21468/SciPostPhys.11.2.041}{{\em
  SciPost Phys.} {\bfseries 11} (2021) 041},
  \href{http://arxiv.org/abs/2101.04759}{{\ttfamily arXiv:2101.04759
  [hep-th]}}.

\bibitem{Peng:2021ago}
H.-H. Peng, J.-J. Zhang, X.-L. Sheng, and Q.~Wang, ``{Ideal Spin Hydrodynamics
  from the Wigner Function Approach},''
  \href{http://dx.doi.org/10.1088/0256-307X/38/11/116701}{{\em Chin. Phys.
  Lett.} {\bfseries 38} no.~11, (2021) 116701},
  \href{http://arxiv.org/abs/2107.00448}{{\ttfamily arXiv:2107.00448
  [hep-th]}}.

\bibitem{Hongo:2021ona}
M.~Hongo, X.-G. Huang, M.~Kaminski, M.~Stephanov, and H.-U. Yee,
  ``{Relativistic spin hydrodynamics with torsion and linear response theory
  for spin relaxation},'' \href{http://dx.doi.org/10.1007/JHEP11(2021)150}{{\em
  JHEP} {\bfseries 11} (2021) 150},
  \href{http://arxiv.org/abs/2107.14231}{{\ttfamily arXiv:2107.14231
  [hep-th]}}.

\bibitem{Hu:2021pwh}
J.~Hu, ``{Relativistic first-order spin hydrodynamics via the Chapman-Enskog
  expansion},'' \href{http://dx.doi.org/10.1103/PhysRevD.105.076009}{{\em Phys.
  Rev. D} {\bfseries 105} no.~7, (2022) 076009},
  \href{http://arxiv.org/abs/2111.03571}{{\ttfamily arXiv:2111.03571
  [hep-ph]}}.

\bibitem{Cartwright:2021qpp}
C.~Cartwright, M.~G. Amano, M.~Kaminski, J.~Noronha, and E.~Speranza,
  ``{Convergence of hydrodynamics in rapidly spinning strongly coupled
  plasma},'' \href{http://arxiv.org/abs/2112.10781}{{\ttfamily arXiv:2112.10781
  [hep-th]}}.

\bibitem{Wang:2021wqq}
D.-L. Wang, X.-Q. Xie, S.~Fang, and S.~Pu, ``{Analytic solutions of
  relativistic dissipative spin hydrodynamics with radial expansion in Gubser
  flow},'' \href{http://arxiv.org/abs/2112.15535}{{\ttfamily arXiv:2112.15535
  [hep-ph]}}.

\bibitem{Hongo:2022izs}
M.~Hongo, X.-G. Huang, M.~Kaminski, M.~Stephanov, and H.-U. Yee, ``{Spin
  relaxation rate for heavy quarks in weakly coupled QCD plasma},''
  \href{http://arxiv.org/abs/2201.12390}{{\ttfamily arXiv:2201.12390
  [hep-th]}}.

\bibitem{Singh:2022ltu}
R.~Singh, M.~Shokri, and S.~M. A.~T. Mehr, ``{Relativistic magnetohydrodynamics
  with spin},'' \href{http://arxiv.org/abs/2202.11504}{{\ttfamily
  arXiv:2202.11504 [hep-ph]}}.

\bibitem{Daher:2022xon}
A.~Daher, A.~Das, W.~Florkowski, and R.~Ryblewski, ``{Equivalence of canonical
  and phenomenological formulations of spin hydrodynamics},''
  \href{http://arxiv.org/abs/2202.12609}{{\ttfamily arXiv:2202.12609
  [nucl-th]}}.

\bibitem{Gallegos:2022jow}
A.~D. Gallegos, U.~Gursoy, and A.~Yarom, ``{Hydrodynamics, spin currents and
  torsion},'' \href{http://arxiv.org/abs/2203.05044}{{\ttfamily
  arXiv:2203.05044 [hep-th]}}.

\bibitem{Weickgenannt:2022zxs}
N.~Weickgenannt, D.~Wagner, E.~Speranza, and D.~Rischke, ``{Relativistic
  second-order dissipative spin hydrodynamics from the method of moments},''
  \href{http://arxiv.org/abs/2203.04766}{{\ttfamily arXiv:2203.04766
  [nucl-th]}}.

\bibitem{Bhadury:2022qxd}
S.~Bhadury, W.~Florkowski, A.~Jaiswal, A.~Kumar, and R.~Ryblewski,
  ``{Relativistic spin-magnetohydrodynamics},''
  \href{http://arxiv.org/abs/2204.01357}{{\ttfamily arXiv:2204.01357
  [nucl-th]}}.

\bibitem{Grozdanov:2016tdf}
S.~Grozdanov, D.~M. Hofman, and N.~Iqbal, ``{Generalized global symmetries and
  dissipative magnetohydrodynamics},''
  \href{http://dx.doi.org/10.1103/PhysRevD.95.096003}{{\em Phys. Rev. D}
  {\bfseries 95} no.~9, (2017) 096003},
  \href{http://arxiv.org/abs/1610.07392}{{\ttfamily arXiv:1610.07392
  [hep-th]}}.

\bibitem{Hongo:2020qpv}
M.~Hongo and K.~Hattori, ``{Revisiting relativistic magnetohydrodynamics from
  quantum electrodynamics},''
  \href{http://dx.doi.org/10.1007/JHEP02(2021)011}{{\em JHEP} {\bfseries 02}
  (2021) 011}, \href{http://arxiv.org/abs/2005.10239}{{\ttfamily
  arXiv:2005.10239 [hep-th]}}.

\bibitem{Grozdanov:2018fic}
S.~Grozdanov, A.~Lucas, and N.~Poovuttikul, ``{Holography and hydrodynamics
  with weakly broken symmetries},''
  \href{http://dx.doi.org/10.1103/PhysRevD.99.086012}{{\em Phys. Rev. D}
  {\bfseries 99} no.~8, (2019) 086012},
  \href{http://arxiv.org/abs/1810.10016}{{\ttfamily arXiv:1810.10016
  [hep-th]}}.

\bibitem{Stephanov:2017ghc}
M.~Stephanov and Y.~Yin, ``{Hydrodynamics with parametric slowing down and
  fluctuations near the critical point},''
  \href{http://dx.doi.org/10.1103/PhysRevD.98.036006}{{\em Phys. Rev. D}
  {\bfseries 98} no.~3, (2018) 036006},
  \href{http://arxiv.org/abs/1712.10305}{{\ttfamily arXiv:1712.10305
  [nucl-th]}}.

\bibitem{Frenkel:1926zz}
J.~Frenkel, ``{Die Elektrodynamik des rotierenden Elektrons},''
  \href{http://dx.doi.org/10.1007/BF01397099}{{\em Z. Phys.} {\bfseries 37}
  (1926) 243--262}.

\bibitem{Huang:2011dc}
X.-G. Huang, A.~Sedrakian, and D.~H. Rischke, ``{Kubo formulae for relativistic
  fluids in strong magnetic fields},''
  \href{http://dx.doi.org/10.1016/j.aop.2011.08.001}{{\em Annals Phys.}
  {\bfseries 326} (2011) 3075--3094},
  \href{http://arxiv.org/abs/1108.0602}{{\ttfamily arXiv:1108.0602
  [astro-ph.HE]}}.

\bibitem{Finazzo:2016mhm}
S.~I. Finazzo, R.~Critelli, R.~Rougemont, and J.~Noronha, ``{Momentum transport
  in strongly coupled anisotropic plasmas in the presence of strong magnetic
  fields},'' \href{http://dx.doi.org/10.1103/PhysRevD.94.054020}{{\em Phys.
  Rev. D} {\bfseries 94} no.~5, (2016) 054020},
  \href{http://arxiv.org/abs/1605.06061}{{\ttfamily arXiv:1605.06061
  [hep-ph]}}. [Erratum: Phys.Rev.D 96, 019903 (2017)].

\bibitem{Hernandez:2017mch}
J.~Hernandez and P.~Kovtun, ``{Relativistic magnetohydrodynamics},''
  \href{http://dx.doi.org/10.1007/JHEP05(2017)001}{{\em JHEP} {\bfseries 05}
  (2017) 001}, \href{http://arxiv.org/abs/1703.08757}{{\ttfamily
  arXiv:1703.08757 [hep-th]}}.

\bibitem{Parodi1970StressTF}
O.~Parodi, ``Stress tensor for a nematic liquid crystal,'' {\em Journal De
  Physique} {\bfseries 31} (1970) 581--584.

\bibitem{RevModPhys.46.617}
M.~J. Stephen and J.~P. Straley, ``Physics of liquid crystals,''
  \href{http://dx.doi.org/10.1103/RevModPhys.46.617}{{\em Rev. Mod. Phys.}
  {\bfseries 46} (Oct, 1974) 617--704}.
  \url{https://link.aps.org/doi/10.1103/RevModPhys.46.617}.

\bibitem{Onsager1931}
L.~Onsager, ``Reciprocal relations in irreversible processes. i.,''
  \href{http://dx.doi.org/10.1103/PhysRev.37.405}{{\em Phys. Rev.} {\bfseries
  37} (1931) 405--426}.

\bibitem{de1984non}
S.~de~Groot and P.~Mazur, {\em Non-equilibrium Thermodynamics}.
\newblock Dover Books on Physics. Dover Publications, 1984.

\bibitem{Landau:Kinetics}
L.~D. Landau and E.~M. Lifshitz, {\em {Physical Kinetics, First Edition}}.
\newblock Butterworth Heinemann, Oxford, UK, 1981.

\bibitem{Green}
M.~S. Green, ``Markoff random processes and the statistical mechanics of
  time‐dependent phenomena. ii. irreversible processes in fluids,''
  \href{http://dx.doi.org/10.1063/1.1740082}{{\em The Journal of Chemical
  Physics} {\bfseries 22} no.~3, (1954) 398--413}.

\bibitem{Nakano}
H.~Nakano, ``A method of calculation of electrical conductivity,''
  \href{http://dx.doi.org/10.1143/PTP.15.77}{{\em Prog. Theor. Phys.}
  {\bfseries 15} no.~1, (1956) 77--79}.

\bibitem{Kubo}
R.~Kubo, ``Statistical-mechanical theory of irreversible processes. i. general
  theory and simple applications to magnetic and conduction problems,''
  \href{http://dx.doi.org/10.1143/JPSJ.12.570}{{\em Journal of the Physical
  Society of Japan} {\bfseries 12} no.~6, (1957) 570--586}.

\bibitem{Kadanoff-Martin1963}
L.~P. Kadanoff and P.~C. Martin, ``Hydrodynamic equations and correlation
  functions,''
  \href{http://dx.doi.org/https://doi.org/10.1016/0003-4916(63)90078-2}{{\em
  Annals of Physics} {\bfseries 24} (1963) 419 -- 469}.

\bibitem{Luttinger:1964zz}
J.~M. Luttinger, ``{Theory of Thermal Transport Coefficients},''
\href{http://dx.doi.org/10.1103/PhysRev.135.A1505}{{\em Phys. Rev.} {\bfseries
  135} (1964) A1505--A1514}.

\bibitem{Belinfante1939}
F.~Belinfante, ``On the spin angular momentum of mesons,''
  \href{http://dx.doi.org/https://doi.org/10.1016/S0031-8914(39)90090-X}{{\em
  Physica} {\bfseries 6} no.~7, (1939) 887 -- 898}.

\bibitem{Belinfante1940}
F.~Belinfante, ``On the current and the density of the electric charge, the
  energy, the linear momentum and the angular momentum of arbitrary fields,''
  \href{http://dx.doi.org/https://doi.org/10.1016/S0031-8914(40)90091-X}{{\em
  Physica} {\bfseries 7} no.~5, (1940) 449 -- 474}.

\bibitem{Rosenfeld1940}
L.~Rosenfeld, ``On the current and the density of the electric charge, the
  energy, the linear momentum and the angular momentum of arbitrary fields,''
  {\em Mem. Acad. Roy. Belg. Cl. Sc.} {\bfseries 18} no.~6, (1940) 1 -- 30.

\bibitem{Becattini:2018duy}
F.~Becattini, W.~Florkowski, and E.~Speranza, ``{Spin tensor and its role in
  non-equilibrium thermodynamics},''
  \href{http://dx.doi.org/10.1016/j.physletb.2018.12.016}{{\em Phys. Lett. B}
  {\bfseries 789} (2019) 419--425},
  \href{http://arxiv.org/abs/1807.10994}{{\ttfamily arXiv:1807.10994
  [hep-th]}}.

\bibitem{Speranza:2020ilk}
E.~Speranza and N.~Weickgenannt, ``{Spin tensor and pseudo-gauges: from nuclear
  collisions to gravitational physics},''
  \href{http://dx.doi.org/10.1140/epja/s10050-021-00455-2}{{\em Eur. Phys. J.
  A} {\bfseries 57} no.~5, (2021) 155},
  \href{http://arxiv.org/abs/2007.00138}{{\ttfamily arXiv:2007.00138
  [nucl-th]}}.

\bibitem{Jensen:2011xb}
K.~Jensen, M.~Kaminski, P.~Kovtun, R.~Meyer, A.~Ritz, and A.~Yarom,
  ``{Parity-Violating Hydrodynamics in 2+1 Dimensions},''
  \href{http://dx.doi.org/10.1007/JHEP05(2012)102}{{\em JHEP} {\bfseries 05}
  (2012) 102}, \href{http://arxiv.org/abs/1112.4498}{{\ttfamily arXiv:1112.4498
  [hep-th]}}.

\end{thebibliography}\endgroup

\end{document}